\documentclass{nic-series}

\begin{document}

\title{
\hspace{8.7cm}{DESY 01-204}\\
\hspace{8cm}{CPT-2001/P.4272}\\
\vspace{0.5cm}
From enemies to friends: chiral symmetry on the lattice}

\author{Pilar Hern\'andez \inst{1} \and Karl Jansen \inst{2} \and Laurent Lellouch \inst{3}}

\institute{CERN, Theory Division,\\
CH-1211 Geneva 23, Switzerland\\
         \email{Pilar.Hernandez@cern.ch}
         \and
         NIC/DESY Zeuthen, \\
         Platanenallee 6, D-15738 Zeuthen, Germany,\\
         \email{Karl.Jansen@desy.de}
         \and
         Centre de Physique Th\'eorique, Case 907, CNRS Luminy,\\
         F-13288 Marseille Cedex 9, France \\
         \email{lellouch@cpt.univ-mrs.fr}
          }

\maketitle

\begin{abstracts}
The physics of strong interactions is invariant under the 
exchange of left-handed and right-handed
quarks, at least in the massless limit. This invariance is reflected 
in the chiral symmetry of quantum chromodynamics. 
Surprisingly, 
it has become clear only recently how
to implement this important symmetry in lattice formulations of quantum field
theories. We will discuss realizations of exact lattice chiral 
symmetry and give an example of the computation of a physical
observable in quantum chromodynamics where chiral symmetry is
important. This calculation is performed by relying 
on finite size scaling methods 
as predicted by chiral perturbation theory. 
\end{abstracts}

\section{Introduction}

Nature as well as physicists like symmetries. The --somewhat amusing-- reason 
for this is that symmetries can be broken. 
A very important concept is the {\em spontaneous breakdown of a symmetry}:
here some symmetry is broken down to a situation with less symmetry when
tuning a parameter (e.g. the temperature or the coupling strength) of the theory
to a critical value.
In the process a non-vanishing vacuum expectation value is developed that
breaks the symmetry of the interaction. 
The process is 
accompanied by the appearance of so-called Goldstone particles 
(spinwaves) that 
are massless. This phenomenon comes under the name of the 
{\em Goldstone theorem}.
An example is the spontaneous magnetization: a metal at high temperature 
is in a symmetric state --
the elementary magnets or spins can point in any direction such that the net 
magnetization vanishes. Decreasing the temperature below some critical
value, a spontaneous magnetization occurs, the spins point into 
a prefered direction and the metal becomes magnetic.

Within the standard model of elementary particle interactions
we know two places where such a spontaneous symmetry breaking 
is supposed to have
occurred: in the electroweak sector spontaneous symmetry breaking
manifests itself in the Higgs phenomenon with the development
of a Higgs field expectation value $\langle \Phi\rangle$, 
giving mass to the elementary particles, and the appearance
of Goldstone particles leading to the W- and Z-bosons. 

In our theory of strong interactions, quantum chromodynamics (QCD), 
it is a chiral symmetry that is assumed to be spontaneously broken.
This symmetry allows for an interchange of left handed and right handed
quarks while leaving physics invariant -- at least when these quarks
are massless. In this case a scalar quark-antiquark $q\bar{q}$ condensate  
is 
developed and  
the Goldstone particles are identified with the light pions that are observed
in nature. 

As stated above the occurrence of spontaneous symmetry breaking 
{\em is an assumption.} The phenomenon is inherently 
non-perturbative
and cannot be addressed with approximative methods like 
perturbation theory.
However, even with numerical simulations it is difficult to test,
whether a certain model exhibits spontaneous symmetry breaking (SSB). 
The reason for this becomes clear when the way to detect 
SSB is considered. 
Let us choose a system that has a finite physical volume $V$ as 
would be required for numerical simulations. Further, we couple
the system to an external magnetic field. Spontaneous symmetry breaking
is tested in 
a double limit, where first the volume of
the system is sent to infinity and then the external magnetic field
is sent to zero. If a non-vanishing magnetization remains, 
spontaneous symmetry breaking is identified. 
Obviously, such a procedure is unfeasible within the approach of
numerical simulations. 

The way out is the use of chiral perturbation theory \cite{gale}. 
In this approach 
chiral symmetry breaking is taken as an assumption with the consequences of 
the appearance of 
non-vanishing field expectation values and Goldstone particles. 
A special situation arises 
when the size of the box becomes comparable to or even smaller than 
the Compton-wavelength 
of the Goldstone particle. Then the corresponding field can be considered
as being uniform and it is possible to set up a systematic expansion 
that starts in the lowest order with an effective lagrangian of 
this constant mode and then taking systematically higher order 
fluctuations into account \cite{xpt}. 

\section{Example of $\Phi^4$-theory}

Let us give an example of the Ginsburg-Landau or $\Phi^4$ theory
with $O(N)$-symmetry in four dimensions. The action of this theory is defined by

\begin{equation}
S= \int d^4 x \frac{1}{2}\left(\partial_\mu\Phi(x)\right)^2 
 + \frac{1}{2} m_0^2 \Phi(x)^2 + \lambda_0 \Phi(x)^4 + j^\alpha \Phi^\alpha(x)
\label{phi4action}
\end{equation}
\newline
with $\Phi$ a $N$-component vector, $m_0$ the bare mass, 
$\lambda_0$ the bare quartic coupling
and $j^\alpha$ a constant external source in direction $\alpha$ of the group
$O(N)$. 

Expectation values of observables are computed through the 
partition function or path integral
in finite volume  $V=L^4$,

\begin{equation} 
\langle \Phi \rangle_{j,V} = \int {\cal D} \Phi \Phi e^{-S}\; .
\end{equation}
Symmetry breaking is detected via a {\em spontaneous magnetization} 
$\langle \Phi \rangle \ne 0$
in the double limit
\begin{equation}
\langle \Phi \rangle = \lim_{j\rightarrow 0} 
         \lim_{V\rightarrow\infty} \langle \Phi \rangle_{j,V}\; .
\end{equation}
Chiral perturbation theory can provide now a prediction for the 
behaviour of $\langle \Phi \rangle_{j,V}$, i.e. the expectation value
in finite volume and at non vanishing external field. In a situation 
when the external field $j$ and $1/V$ are very small,
$\langle\Phi \rangle_{j,V}$ is given in terms of the expectation value 
of $\langle\Phi \rangle$ at $j=0$ and $V=\infty$: 

\begin{equation}
\langle \Phi \rangle_{j,V} = f(u) 
\label{eq:phi}
\end{equation}
\noindent where $u$ is a scaling variable
\begin{equation}
u=\langle \Phi\rangle jV
\end{equation}
that
contains the infinite volume and $j=0$
spontaneous
magnetization $\langle \Phi \rangle$. 

Let us give for completeness 
the function $f(u)$:
\begin{equation}
f(u)=\frac{u^2 \eta(u)}{jV}\;\;\; \mathrm{with}\;\;\; 
\eta(u) = \frac{1}{u}\frac{I_2(u)}{I_1(u)}
\end{equation}
with $I_1$, $I_2$ modified Bessel functions. The important point 
here is that these functions only depend on a single scaling variable
$u$ and hence on the quantity of interest, the magnetization 
$\langle \Phi \rangle$.

In an already quite old work \cite{fss:karl91} the prediction of chiral perturbation
theory, eq.~(\ref{eq:phi}),
was confronted with numerical data. In this test the data obtained
through numerical simulations were described very well by the
theoretical formulae of chiral perturbation theory and  
the method of chiral perturbation theory turned out
to be very fruitful, at least in this example of a relatively simple 
model. The lesson that is to be learnt from this example is that 
finite size effects can actually be used to determine 
properties of an infinitely
large system. In this sense, the finite volume system can 
be regarded as a probe of the target theory in infinite volume. 
This gives an entirely new perspective
on finite size effects:
instead of being afraid 
of them, they can be used to determine 
important physical information. 

\section{The example of quantum chromodynamics}

Chiral perturbation theory as well as the lattice method was developed
for quantum chromodynamics in order to deepen our understanding of the 
strong interactions. 
Still, until relatively recently, both approaches could not 
really come together, at least in 
the region of very small quark masses where chiral symmetry starts to get
restored. The reason was that the lattice seemed to be lacking the
concept of chiral symmetry and for many years the infamous 
Nielson-Ninomiya theorem \cite{nn} was telling us that it would 
even be impossible
to implement chiral symmetry in a consistent way on a lattice. 

The situation only changed a few years ago, when an old work by 
Ginsparg and Wilson~\cite{chiral:GWR} was rediscovered \cite{Hasenfratz:1998ri}. 
The Ginsparg-Wilson paper contained actually 
a clue for answering the
problem of chiral fermions on the lattice.  
The interaction of the fermions is described by some particular
operator, the Dirac operator, the details of which should not
be discussed here. 
Now, in the continuum theory this Dirac operator 
anticommutes with a certain 4-dimensional matrix 
$\gamma_5=\mathrm{diag}(1,1,-1,-1)$. On a lattice with non-vanishing 
lattice spacing $a$, such an anti-commutation property cannot be
demanded. If the anticommutation property is insisted on, the fermion 
spectrum of the lattice theory does not correspond to the
one of the target continuum theory. 

The suggestion of Ginsparg and Wilson was to replace the 
anticommutation condition 
by a relation (now known as the Ginsparg-Wilson (GW) relation)
for a lattice Dirac operator $D$:
\begin{equation}
\gamma_5D + D\gamma_5 = aD\gamma_5D\; .
\label{gwrelation}
\end{equation}
Clearly, in the limit that the lattice spacing vanishes
the usual anti-commutation relation of the continuum theory is recovered. 

The fact that renders the relation eq.~(\ref{gwrelation}) conceptually 
extremely fruitful is that it implies an 
exact chiral symmetry on the lattice even if the value of the lattice spacing 
does not vanish \cite{Martin_exact}.                    
The notion of a chiral symmetry on the lattice is a conceptual 
breakthrough and renders 
the lattice theory 
in many respects to behave like its continuum counterpart with far reaching 
consequences. 

However, as nice as the theoretical progress that followed the 
rediscovery of the Ginsparg-Wilson relation was as much of a challenge 
are realizations of operators $D$ that satisfy the Ginsparg-Wilson relation
(see \cite{nieder,blum,pilar} for reviews). 
Let us give a particular example for such a solution as found by 
H.~Neuberger \cite{chiral:ovlp} from the overlap formalism \cite{ovlp},
based on the pioneering work of D. Kaplan \cite{kaplan}. 
To this end, we first consider the standard 
Wilson Dirac operator on the lattice:

\begin{equation}
  D_{\rm w}=\frac{1}{2}\left\{\gamma_{\mu}(\nabla_{\mu}^{*}+\nabla_{\mu})
  -a\nabla_{\mu}^{*}\nabla_{\mu}\right\}
\end{equation}
with $\nabla_{\mu}$, $\nabla_{\mu}^{*}$  
the lattice forward, backward derivatives, i.e. nearest neighbour differences,
acting on a field $\Phi(x)$
\begin{eqnarray}
\nabla_{\mu}\Phi(x) & = & \Phi(x+\mu) - \Phi(x) \nonumber \\
\nabla^{*}_{\mu}\Phi(x) & = & \Phi(x) - \Phi(x+\mu)\; . \nonumber 
\end{eqnarray}
We then define 
\begin{equation}
A=1+s - D_{\rm w}
\end{equation} 
with $0 < s < 1$ a tunable parameter. 
Then Neuberger's operator
$D_{\rm N}$ with mass $m$ is given by

\begin{equation}
  D_{\rm N}=\left\{1 - \frac{m}{2(1+ s)} D_{\rm N}^{(0)}+m\right\}
\end{equation}
where
\begin{equation}
  D_{\rm N}^{(0)}=(1+ s)\left[ 1-A(A^{\dagger}A)^{-1/2} \right]\; .
\end{equation}

What is important here in the definition of Neuberger's operator 
is the appearance of the square root of the operator 
$A^{\dagger}A$. 
This means that $D_{\rm N}$ connects all points of the lattice
with each other. Note, however, that despite this the operator
is a local operator in the field theoretical sense \cite{hjl}. 
In practice, the operator $A^{\dagger}A$ is represented by a matrix that is,
unfortunately, very large. Having a physical volume of size $L^4$, the 
number of sites is $N^4=(L/a)^4$. As the internal number of degrees of freedom
per lattice point 
is $12$, we end up with $A$ being a $(12N^4)\otimes(12N^4)$ complex matrix 
with $N$ typically in the range $10<N<30$ for present days simulations. 
Hence we have to construct the square root of a very large matrix. 
What is worse, to compute relevant physical observables, we don't
need the operator $D_{\rm N}$ itself but its inverse. Such an inverse
is constructed by iterative methods like the conjugate gradient 
algorithm and 
its relatives \cite{numrec}. 

The square root can be constructed by a polynomial expansion, 
normally based on a Chebyshev approximation, or by rational
approximations. Both methods give comparable performances in 
practice. 
The convergence of the approximation to the square root is determined 
by the condition number of the (positive definite) matrix 
$A^\dagger A$. 
When the matrix $A^\dagger A$ is normalized such that the largest 
eigenvalue is one, 
the condition number is given by the inverse of the lowest eigenvalue. 
We show in figure~1 the low-lying eigenvalues of $A^\dagger A$ for different
values of $\beta$ where the parameter $\beta$ is inversely proportional to the
coupling strength of the theory.

\begin{figure}[t]
\begin{center}
\epsfxsize=12cm
\epsfysize=12cm
\epsfbox{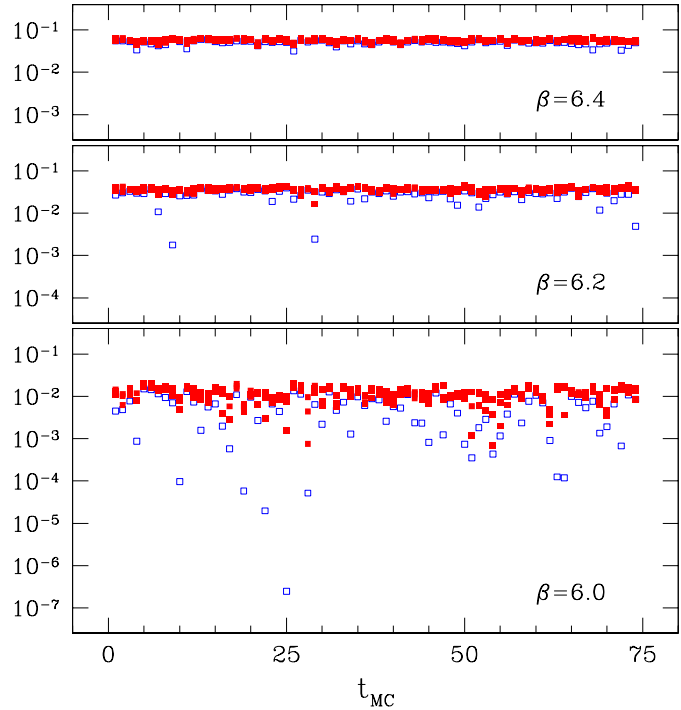}
\vspace{-0.7cm}
\caption{\label{ei_figure}
The low-lying end of the eigenvalue spectrum of the matrix 
$A^\dagger A$. 
}
\end{center}
\vspace{-0.2cm}
\end{figure}

It can clearly be observed that very small eigenvalues can occur resulting in 
large condition numbers. In such situations the convergence of the approximations
chosen can be rather slow and special tricks have to be implemented to
accelerate the convergence. The most fruitful improvement is to treat
a part of the low-lying end of the spectrum exactly by projecting this part
out of the matrix $A^\dagger A$ \cite{cond:SCRI1,technical}. 
Further improvements can be implemented, examples of which 
are discussed in ref.~\cite{technical}.

Despite all these technical improvements it is found that
a typical value for the degree of a polynomial is
$O(100)$ and a typical value for the number of iterations to
compute the inverse of $D_\mathrm{N}$ is again $O(100)$. 
Since in each iteration to compute  
$D_\mathrm{N}^{-1}$ the Chebyshev polynomial has to be 
evaluated, this means that for 
a value of a physical observable on a single configuration 
ten thousand applications
of a huge matrix on a vector has to be performed. To compute 
the expectation value of some observable, this observables has to
be averaged over many gluonic configurations.  
Clearly, this results in a very demanding computational
effort and gives rise to a numerical challenge well 
suited for NIC.

The only thing that helps a lot in this problem is that 
the matrix $A$ is sparse. Only the diagonal and a few 
sub-diagonals are actually filled, a fact that finally makes
the problem manageable -- although, still a very large amount
of computer resources are needed to tackle it.              

\section{The scalar condensate}

The existence of an exact lattice chiral symmetry allows 
the use of finite size effects to test for SSB in QCD as 
in the case of the $\Phi^4$-theory. Using a chiral invariant formulation 
of lattice QCD, it is possible
to reach the region of very small quark masses where it is to be expected
that chiral symmetry starts to get restored. 

The ``magnetization'' in the case of quantum chromodynamics is the 
condensate of a quark-antiquark state $\Sigma=\langle \psi\bar{\psi}\rangle$. 
The role of the external magnetic field is played by the quark mass $m_q$.
We have developed a
fully parallelized code with all technical improvements 
implemented. This allowed us to compute
the scalar condensate as a function of the quark mass at several volumes. 
A (standard) caveat here is the fact that all computations are done
in the so-called quenched approximation where all internal quark 
loops are neglected.
In QCD there is a peculiarity: the field configurations can have 
topological properties, characterized by the so-called 
topological charge which can be measured --unambiguously-- through the
number of zero modes of the operator $D_\mathrm{N}$. In fact, the formulae from
quenched chiral perturbation theory are parametrized by the topological
charge and it is hence very important to be able to identify 
the topological charge of the gauge field configurations. 
Without the special properties of lattice Dirac operators that satisfy 
the Ginsparg-Wilson relation such an identification would be very difficult.

Let us give the complete theoretical formula from quenched 
chiral perturbation theory in lowest order:
\begin{equation}
\Sigma_{\nu=\pm 1}(m_q,V) = \Sigma\; z \; [ I_{\nu}(z) K_{\nu}(z) + I_{\nu+1}(z)
K_{\nu-1}(z)] + C\cdot m_q/a^2\; .
\label{chipt}
\end{equation}
The only important thing to notice here is that this relatively involved
combination of Bessel functions do, as in the case of the $\Phi^4$-theory,
 only depend on one scaling variable
\begin{equation}
z=\Sigma m_qV
\end{equation}
that contains the quantity of interest, namely the infinite volume,
chiral limit scalar condensate $\Sigma$. The additional term 
$C\cdot m_q/a^2$ is
a power divergence that comes from the renormalization properties
of the theory. We will not discuss this field theoretical aspect here
but just notice that this term has to be included in the fit.

\begin{figure}[t]
\begin{center}
\epsfxsize=12cm
\epsfysize=12cm
\epsfbox{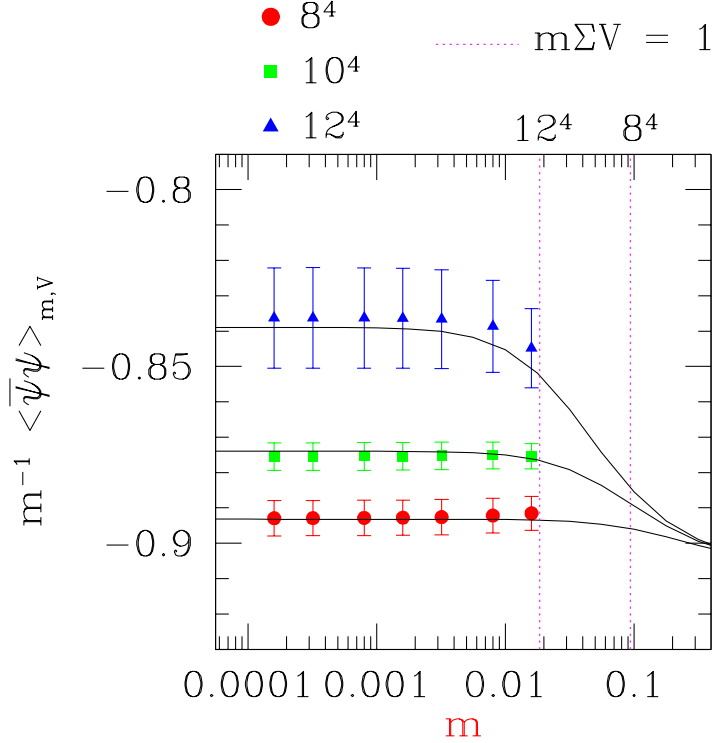}
\vspace{-0.7cm}
\caption{\label{cond_figure}
The scalar condensate computed on lattices of various size as a
function of the quark mass. The solid lines are 2-parameter ($\Sigma$ and the 
constant $C$ of eq.~(\ref{chipt})) fits according to the
prediction of chiral perturbation theory. 
}
\end{center}
\vspace{-0.2cm}
\end{figure}

In figure~2, we show the result of our numerical computation of the 
scalar condensate~\cite{cond:paperI} 
in a fixed topological charge sector $|\nu|=1$ as a 
function of the quark mass at several volumes. The solid line is a fit to
the prediction of chiral perturbation theory, eq.(\ref{chipt}). We find that 
the simulation data are described by this prediction very well. 
This means that we find evidence for the basic assumption on which
the theoretical prediction relies: the appearance of spontaneous
chiral symmetry breaking in (quenched) QCD.

We want to remark that this work that has been performed at NIC was the first
of this kind. The project consumed 1400 CPU hours on a typical 
distribution of the lattice on 128 nodes. 
After this work, a number of other groups repeated such an
analysis \cite{cond:DeGrand,cond:giusti,cond:hase} and it was reassuring 
to observe that very consistent results 
were found. In a subsequel work \cite{cond:paperIII,cond:paperIV} 
we developed also a quite general method
for renormalizing the value of the bare scalar condensate as extracted
from the finite size scaling analysis performed here. 

\section{Conclusion}

In this contribution we have demonstrated that by a combination of 
theoretical ideas, improved numerical methods and the use of 
powerful supercomputer platforms it is possible to test basic
properties of field theories. Of particular interest was the 
question of whether the phenomenon of spontaneous symmetry breaking
does occur in certain field theories important in elementary particle
physics. The phenomenon of SSB leads to far reaching consequences in theories
like QCD or the scalar sector of the electroweak interactions. 
In the work performed here, we found strong 
evidence for the appearance of spontaneous chiral
symmetry breaking in quenched lattice QCD.  
This conclusion is the result 
of the fact that two
theoretical concepts, the lattice approach to quantum field theories
and chiral symmetry, met finally -- and that enemies became friends.

We are very much indebted to Hartmut Wittig for numerous discussions,
suggestions and finally participating in the project at the stage 
when the non-perturbative renormalization of the scalar
condensate was computed. 
L.L. thanks the INT at the University of Washington for its hospitality
and the DOE for partial support during the completion of this
work. This work was supported in part by the EU TMR program under
contract FMRX-CT98-0169 and the EU HP program under contract
HPRN-CT-2000-00145.

\end{document}